\documentclass[prl,final,twocolumn,showpacs,showkeys]{revtex4}
\usepackage{amsmath}
\usepackage{graphicx}
\usepackage{amsfonts}
\usepackage{amssymb}
\usepackage{subfigure}
\begin{document}

\title{Connecting the discrete and continuous-time quantum walks}

\author{Frederick W. Strauch}
\email[Electronic address: ]{frederick.strauch@nist.gov}
\affiliation{National Institute of Standards and Technology, Gaithersburg, Maryland 20899-8423, USA}

\date{\today}

\begin{abstract}
Recently, quantized versions of random walks have been explored as effective elements for quantum algorithms.  In the simplest case of one dimension, the theory has remained divided into the discrete-time quantum walk and the continuous-time quantum walk.  Though the properties of these two walks have shown similarities, it has remained an open problem to find the exact relation between the two.  The precise connection of these two processes, both quantally and classically, is presented.  Extension to higher dimensions is also discussed. 
\end{abstract} 
\maketitle

Continuous-time quantum walks (CTQW) were introduced by Farhi and Guttman \cite{Farhi98} as generalizations of diffusion-type differential equations, in which probability is replaced by a complex amplitude and Markovian dynamics is replaced by unitary dynamics.  Their motivation was to explore whether, in a given framework, coherent quantum processes could show dramatic differences from classical random walks.  This turned out to be the case, for Childs {\it{et al.}} demonstrated a graph problem that could be solved using the CTQW exponentially faster than not just classical random walks but all classical methods \cite{Childs2003}.  In one dimension, the CTQW is simply the finite-difference Schr{\"o}dinger equation \cite{Childs2002}
\begin{equation}
i\partial_t \psi(n,t) = - \gamma [ \psi(n+1,t)-2 \psi(n,t) + \psi(n-1,t) ],
\label{ctqw}
\end{equation}
where $\psi(n,t)$ is a complex amplitude at the (continuous) time $t$ and (discrete) lattice position $n$.  

The discrete-time quantum walk (DTQW), introduced by Aharonov {\textit{et al.}} \cite{Aharonov93} and independently by Meyer \cite{Meyer96}, is a discrete unitary mapping such as
\begin{equation}
\begin{array}{l}
\psi_R(n,\tau+1) = \cos \theta \psi_R(n-1,\tau) - i \sin \theta \psi_L(n-1,\tau) \\ 
\psi_L(n,\tau+1) = \cos \theta \psi_L(n+1,\tau) - i \sin \theta \psi_R(n+1,\tau),
\end{array}
\label{dtqw}
\end{equation}
where $\psi_R(n,\tau)$ and $\psi_L(n,\tau)$ are complex amplitudes at the (discrete) time $\tau$ and (discrete) lattice position $n$, and the labels $R$ and $L$ indicate an additional degree of freedom, often taken as the state of a coin which tells the walker (located at position $n$) which way to step.  This discrete dynamics has a rich mathematical structure that is quite foreign to the CTQW, and has been the subject of extensive theoretical investigation.  In particular, there has been significant extension of the DTQW to include decoherence \cite{Brun2003}, quantum chaotic \cite{Wojcik2003} and quasiperiodic effects \cite{Ribeiro2004}.  In addition, certain continuum limits have been used to connect the DTQW to more familiar wavelike propagation \cite{Knight2003,Blanchard2004,Strauch2006}.  Despite this large body of work, the relation of these two quantum walks remains an open problem.

This problem is truly fundamental for quantum computation, for at least two reasons.  First, it is quite unnatural to have two distinct ways to quantize classical diffusion.  Determining whether quantum mechanics speeds up a classical process is difficult enough, but even moreso if there is no unique quantization.  For the processes considered here, the coin degree of freedom appears unnecessary, and indeed there is a perfectly reasonable discrete-time quantum process that can be implemented {\textit{without}} a coin \cite{Patel2005}---this will be discussed below.  Second, the spreading properties of the two quantum walks are quite similar \cite{Strauch2006}.  From an initially localized state, both evolutions generate a probability distribution that is nearly constant save for two peaks at $\pm c t$ (here $c=2\gamma$ for (\ref{ctqw}) and $c = \cos \theta$ for (\ref{dtqw})), decaying to zero thereafter (see Figure 1).  Both have standard deviations of position that grow linearly in time, quadratically faster than classical diffusion.  These similarities suggest that, besides the fact that both are unitary quantum processes, there should be some underlying connection between the two walks.  Nevertheless, the precise relationship has remained elusive.  In particular, no one has demonstrated how to get (\ref{ctqw}) from (\ref{dtqw}) by some limiting process.  

\begin{figure*}
\centering
\subfigure[ DTQW]{\includegraphics[width=2 in]{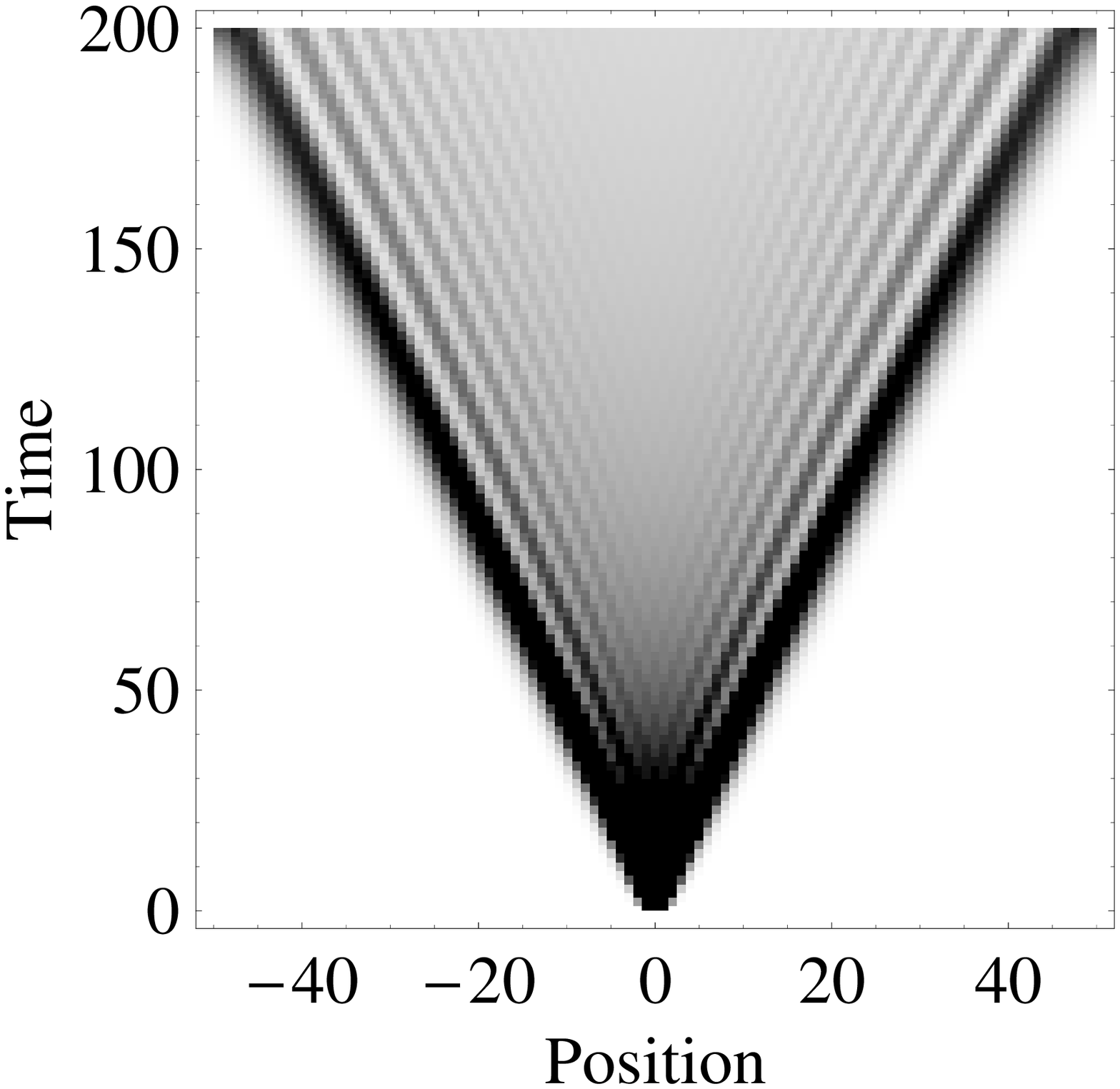}}
\subfigure[{Continuous-time limit of DTQW}]{\includegraphics[width = 2 in]{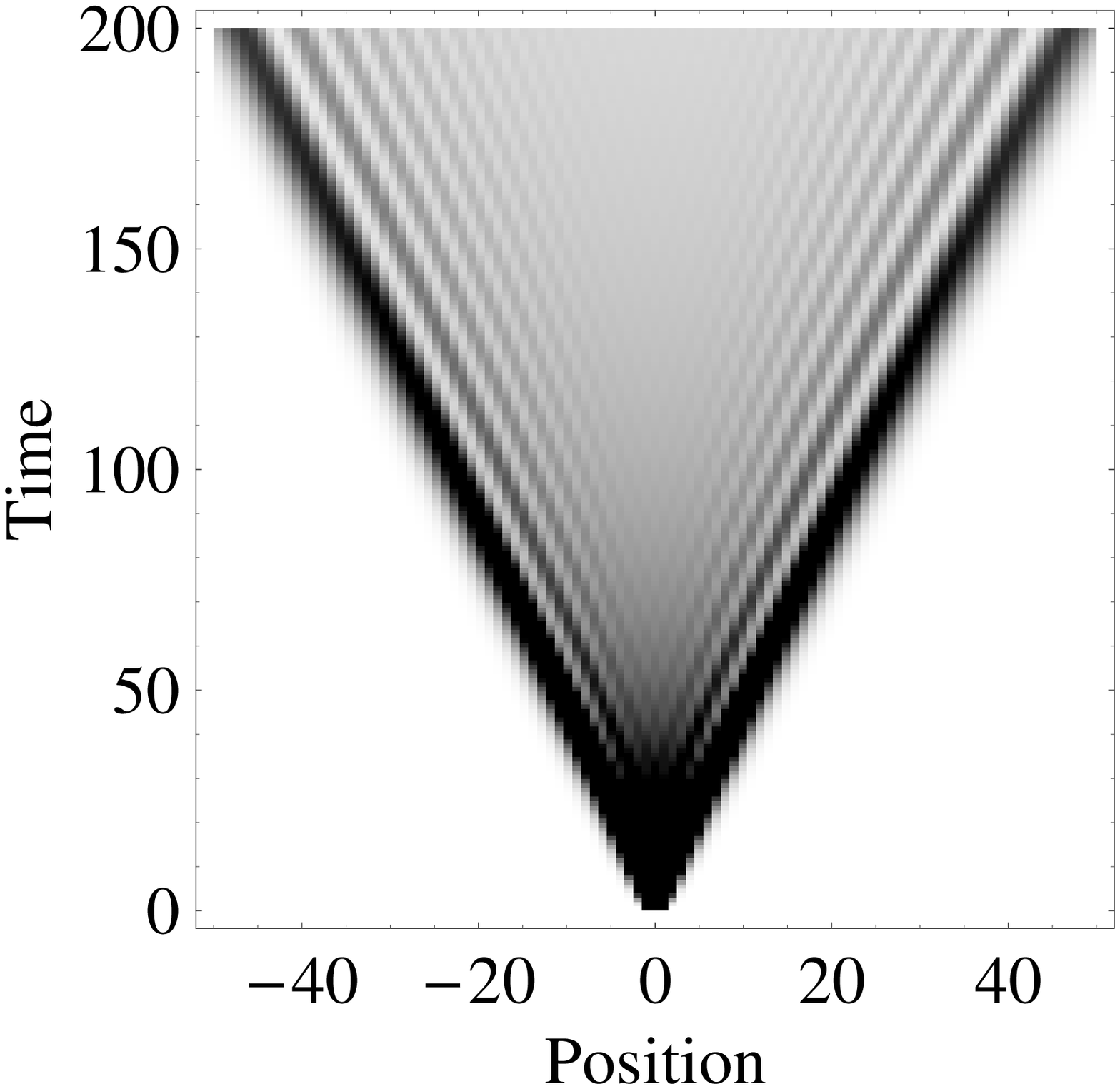}}
\subfigure[ CTQW]{\includegraphics[width = 2 in]{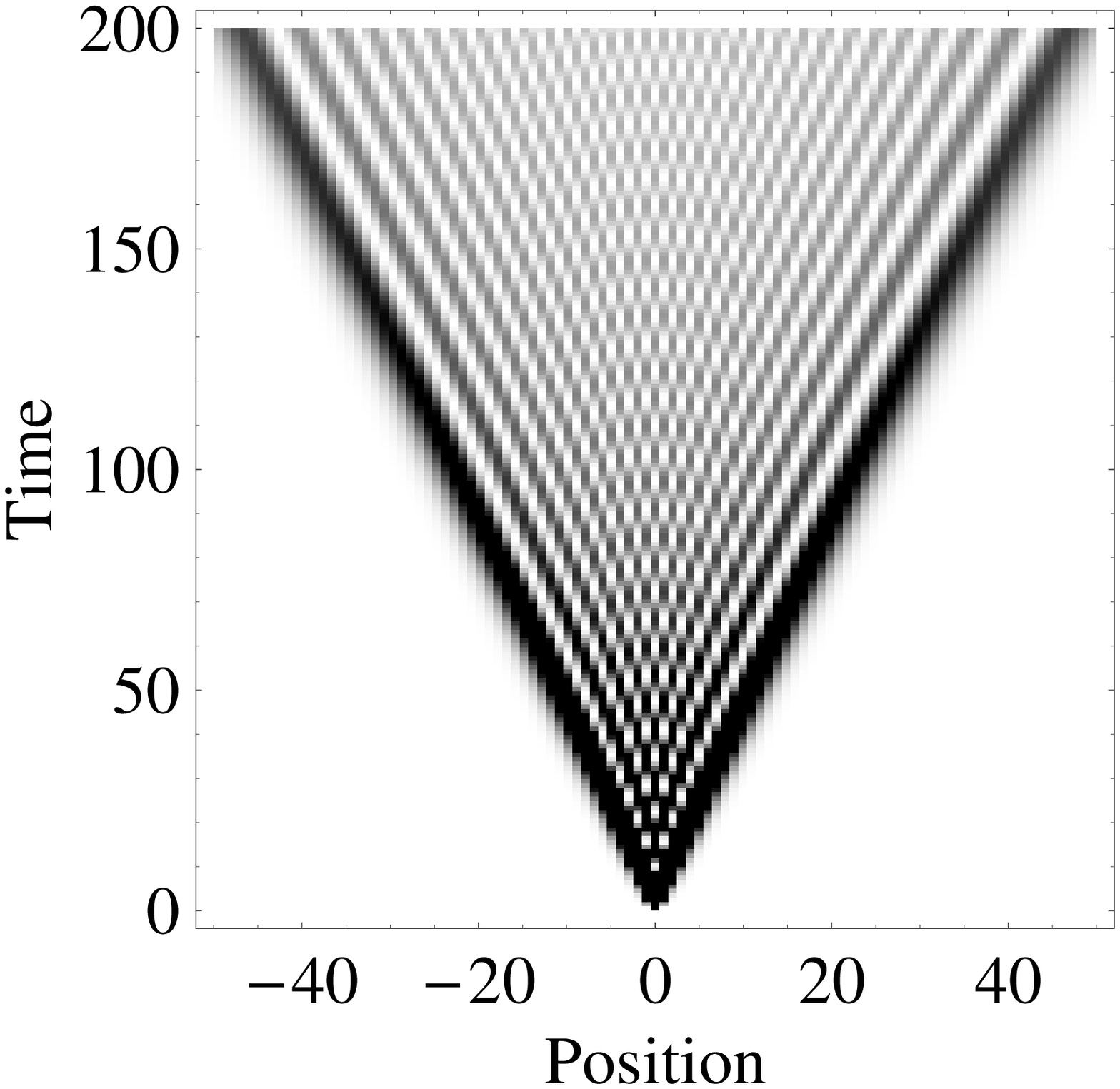}}
\caption{Evolution of (a) the DTQW with $\cos \theta = 1/4$, (b) the $\theta \to \pi/2$ continuous-time limit of the DTQW, with $\gamma = 1/8$, and (c) the CTQW with $\gamma = 1/8$.  The probability density $\rho(n,t) = \Psi(n,t)^{\dagger} \Psi(n,t)$ is shown, where the initial conditions for (a) and (b) are given by (\ref{initcond}), while for (c) $\psi(n,0) = \delta_{n,0}$.}
\label{onlyfig}
\end{figure*}

Aside from certain approximations verified numerically \cite{Romanelli2004L,Strauch2006}, the closest previous connection of these two walks is the weak limit theorems for the probability density due to Konno and others \cite{Konno2002}.  Specifically, letting $n \sim x$, where $x$ is considered a continuous variable, the long-time limit of the CTQW probability density is
\begin{equation}
P_{CTQW}(x,t) \approx \frac{1}{\pi \sqrt{(2 \gamma t)^2 - x^2}}
\label{xqw1}
\end{equation}
where $-2\gamma t < x < 2 \gamma t$, while the long-time limit of the DTQW probability density is
\begin{equation}
P_{DTQW}(x,\tau) \approx \frac{\sin \theta}{\pi (1-x^2 \tau^{-2}) \sqrt{ (\cos \theta \tau)^2 - x^2}}.
\label{xqw2}
\end{equation} 
with $-\cos \theta \tau < x < \cos \theta \tau$.  Comparing these two expressions, one might be led to consider the limit $\tau \to \infty$, $\theta \to \pi/2$, such that $\cos \theta \tau \to 2 \gamma t$.  This certainly maps (\ref{xqw2}) to (\ref{xqw1}), but what about (\ref{ctqw}) and (\ref{dtqw})?  Here I will show that this limit does indeed map the DTQW to the CTQW, has a direct parallel with the relevant classical random walks, and can be extended to higher-dimensional walks.

The most well known limit of the DTQW \cite{Meyer96}, with $\theta \to 0$, was introduced by Feynman in 1946 to construct a path integral for the propagator of the Dirac equation \cite{Schweber86}.  In Feynman's picture, a particle zig-zags at the speed-of-light across a space-time lattice, flipping its chirality from left to right with an infinitesimal probability each time-step.  The resulting dynamics, in the continuum limit \footnote{{The precise limit requires introducing space-time lattice spacings of $\Delta x$ and $\Delta t$.  Then, one lets $\theta = m c^2 \Delta t / \hbar$, $\Delta x = c \Delta t$, taking the limit $\Delta t \to 0$ with the continuum coordinates $x = n \Delta x, t = \tau \Delta t$ held constant.}}, is the Dirac equation, with the flipping rate determined by the mass of the particle.   

The limit considered here is, at first sight, quite puzzling.  With $\theta \to \pi/2$, the DTQW describes a particle flipping its chirality with nearly {\textit{unit}} probability each time-step.  Such a particle should not move at all, and indeed the maximum group velocity is $\cos \theta$ \cite{Strauch2006}, which goes to zero in the same limit.  However, one must also go from discrete to continuous time---only by taking these two limits together does the DTQW become the CTQW.  

First, it is convenient to work in momentum space by introducing the Fourier transform
\begin{equation}
\Psi(n,\tau) = \left(\begin{array}{c} \psi_R(n,\tau) \\ \psi_L(n,\tau) \end{array} \right) = \frac{1}{2\pi} \int_{-\pi}^{\pi} dk \left(\begin{array}{c} \phi_R(k,\tau) \\ \phi_L(k,\tau) \end{array} \right) e^{i n k}.
\label{fourier}
\end{equation}
Then, the DTQW after $\tau$ steps is the unitary mapping
\begin{equation}
\left(\begin{array}{c} \phi_R(k,\tau) \\ \phi_L(k,\tau) \end{array} \right) = U (\tau) 
\left(\begin{array}{c} \phi_R(k,0) \\ \phi_L(k,0) \end{array} \right),
\label{fourier1}
\end{equation}
where $U(\tau) = U^{\tau}$ with the single-step propagation matrix 
\begin{equation}
U = e^{-i k \sigma_z} e^{-i \theta \sigma_x} =\left(\begin{array}{cc} e^{-ik} \cos \theta & -i e^{-i k} \sin \theta \\
                        -i e^{i k} \sin \theta & e^{i k} \cos \theta \end{array} \right),
\label{fourier2}
\end{equation}
and I have introduced the Pauli matrices $\sigma_x, \sigma_y, \sigma_z$.

The key step is to set $\theta = \pi/2 - \delta$, where $\delta \ll 1$.  Then, using $e^{-i\theta \sigma_x} = -i \sigma_x e^{i \delta \sigma_x}$,  I compute $U^2$:
\begin{equation}
\begin{array}{ll}
U^2 &= (-i)^2 e^{-i k \sigma_z} \sigma_x e^{i \delta \sigma_x} e^{-i k \sigma_z} \sigma_x e^{i \delta \sigma_x} \\
    &= (-i)^2 e^{-i k \sigma_z} e^{i \delta \sigma_x} e^{i k \sigma_z} e^{i \delta \sigma_x} \\
    &= (-i)^2 \exp [i \delta( \sigma_x \cos 2k + \sigma_y \sin 2k)] e^{i \delta \sigma_x} \\
    &= (-i)^2 \exp [i \delta (\sigma_x (1+\cos 2k) + \sigma_y \sin 2k)] + O(\delta^2) \\
    &= (-i)^2 \exp [i \delta 2 \cos k (\sigma_x \cos k + \sigma_y \sin k)] + O(\delta^2),
\end{array}
\label{fourier3}
\end{equation}
where I have used properties of the Pauli matrices in the second and third lines, the Baker-Campbell-Hausdorff theorem in the fourth, and trigonometric identities in the last.  Now, applying $U^2$ for $\tau/2$ times, and taking the limit $\delta \to 0$, $\tau \delta \to 2 \gamma t$, I find
\begin{equation}
\begin{array}{ll}
U(\tau) &= (-i)^{\tau} \exp [i 2 \gamma t \cos k (\sigma_x \cos k + \sigma_y \sin k)] \\
& \equiv \exp(-i \Phi) \exp(-i H t),
\end{array}
\label{fourier4}
\end{equation}
where I have defined $\Phi = \tau \pi/2$ and 
\begin{equation}
H = -2 \gamma \cos k (\sigma_x \cos k + \sigma_y \sin k).
\label{fourier5}
\end{equation}

Using the result (\ref{fourier4}) in (\ref{fourier1}) shows that aside from the unimportant overall phase $\Phi$, this limit of the DTQW describes continuous time evolution with a Hamiltonian given by (\ref{fourier5}).  The corresponding Schr{\"o}dinger equation $i \partial_t \Psi = H \Psi$, found by the Fourier transform (\ref{fourier}), is
\begin{equation}
\begin{array}{ll}
i \partial_t \psi_R(n,t) &= -\gamma [ \psi_L(n,t) + \psi_L(n-2,t) ] \\
i \partial_t \psi_L(n,t) &= -\gamma [ \psi_R(n,t) + \psi_R(n+2,t)],
\end{array}
\label{ctqwlimit4}
\end{equation}
where here and in the following the amplitudes have been trivially redefined, e.g. $\psi_R(n,\tau) \to \psi_R(n,t)$.

The reduction to the CTQW is nearly complete.  To make the connection fully transparent, first observe that a general solution of (\ref{ctqwlimit4}) can be split into two terms
\begin{equation}
\Psi(n,t) = e^{i 2 \gamma t} \Psi_{+}(n,t) + e^{-i 2 \gamma t}\Psi_{-}(n,t)
\label{ctqwlimit5}
\end{equation}
where
\begin{equation}
\Psi_{+}(n,t) = \frac{1}{2} e^{-i 2 \gamma t} \left( \begin{array}{c} \psi_R(n,t) + \psi_L(n-1,t) \\ \psi_L(n,t) + \psi_R(n+1,t) \end{array} \right)
\end{equation}
and
\begin{equation}
\Psi_{-}(n,t) = \frac{1}{2} e^{i 2 \gamma t} \left( \begin{array}{c} \psi_R(n,t)-\psi_L(n-1,t) \\ \psi_L(n,t) - \psi_R(n+1,t) \end{array} \right).
\end{equation}
By direct substitution one finds
\begin{equation}
\begin{array}{l}
i\partial_t \Psi_{\pm}(n,t) \\
\qquad = \mp \gamma [ \Psi_{\pm}(n+1,t) -2 \Psi_{\pm}(n,t) + \Psi_{\pm}(n-1,t) ].
\end{array}
\label{ctdtqw}
\end{equation}
Thus, the continuous-time limit with $\theta \to \pi/2$ of the DTQW is really equivalent to {\it{two}} copies of the CTQW.  As a consequence, there still remains interference associated with the combination of the two components $\Psi_{\pm}(n,t)$ in (\ref{ctqwlimit5}), which oscillates in time.  This can be eliminated by requiring $\psi_R(n,t)=\psi_L(n-1,t)$, so that $\Psi_-(n,t)=0$ \footnote{{The condition $\Psi_-(n,t)=0$ need only be enforced at time $t=0$.  Linearity of the differential equation (\ref{ctqwlimit5}) ensures that it will then be satisfied for all $t$}}; an equivalent simplification was previously found by projecting the initial state onto the ``positive-energy'' subspace of $U$ \cite{Strauch2006}.  

The simplest initial condition with $\Psi_-(n,t)=0$ and localized symmetrically about $n=0$ takes the form
\begin{equation}
\left(\begin{array}{l} \psi_R(n,0) \\ \psi_L(n,0) \end{array}\right) = \frac{1}{2} \left(\begin{array}{l} \delta_{n,0} + \delta_{n,1} \\ \delta_{n,-1} + \delta_{n,0} \end{array}\right).
\label{initcond}
\end{equation}
The subsequent evolution, using a numerical evaluation of the DTQW (\ref{dtqw}) is shown in Fig. 1(a), using a reasonably small value of $\cos \theta$.  The solution of the continuous-time limit (\ref{ctdtqw}) can be found analytically in terms of the regular Bessel function \cite{Childs2002,Strauch2006}:
\begin{equation}
\left(\begin{array}{l} \psi_R(n,t) \\ \psi_L(n,t) \end{array}\right) = \frac{1}{2} i^n \left(\begin{array}{l} J_n(2\gamma t) - i J_{n-1}(2 \gamma t) \\ J_n(2 \gamma t) + i J_{n+1} (2 \gamma t) \end{array} \right),
\label{dtqwbessel}
\end{equation}
whose probability density is shown in Fig. 1(b).  Note that the initial condition (\ref{initcond}) differs from most previous studies of the DTQW: (i) it has support on $n=0$ and $n= \pm 1$ and (ii) it is entangled.  Both are responsible for the slightly blurred appearance in comparison to the CTQW $\psi(n,t) = e^{-2 i \gamma t} i^n J_n(2 \gamma t)$ shown in Fig. 1(c).  Nevertheless, the visual agreement between all three is quite clear, and the analytical result (\ref{dtqwbessel}) shows that the approximation found previously \cite{Strauch2006} is exact in the $\theta \to \pi/2$ limit.

This result---the fact that the DTQW limits to the CTQW---agrees with, from an entirely new perspective, the interpretation of the quantum walk as a simple interference process \cite{Knight2003} in which the coin degree of freedom is, at least in this simple case, irrelevant to the speedup found in quantum walk algorithms.  In this context, it has been shown that a discrete-time evolution can be constructed without a coin \cite{Patel2005}.  Here, one takes the Hamiltonian that generates the CTQW and splits it into ``even'' and ``odd'' terms $H = H^{even} + H^{odd}$ \footnote{{For example, with $H_{n,m} = 2 \delta_{n,m} - \delta_{n+1,m} - \delta_{n-1,m}$ let $H_{n,m}^{even} = \delta_{n,m} - \frac{1}{2}(1+(-1)^n) \delta_{n+1,m} - \frac{1}{2}(1-(-1)^n) \delta_{n-1,m}$ and $H_{n,m}^{odd} =  \delta_{n,m} - \frac{1}{2}(1-(-1)^n) \delta_{n+1,m} - \frac{1}{2}(1+(-1)^n) \delta_{n-1,m}$.}}. Such a splitting is favorable for computation (classical \cite{Richardson91} or quantum \cite{Boghosian98b}), since $H^{even}$ and $H^{odd}$ are both block-diagonal matrices, with each block a simple 2 by 2 matrix.  Then, one constructs the unitary operator $U(\theta_1,\theta_2) = \exp(i\theta_2 H^{odd}) \exp(i \theta_1 H^{even})$ (the case $\theta_1 = \theta_2 = \pi/4$ was considered in \cite{Patel2005}).  Though this operator trivially becomes equivalent (in the continuous-time limit) to the CTQW for $\theta_1 = \theta_2 \to 0$, the limit found here is quite distinct.  By using the even-odd splitting of the lattice, one can formally reintroduce the coin degree of freedom and recover the DTQW (\ref{dtqw}) with $\theta_1 = \theta-\pi/2$ and $\theta_2 = \pi/2$.  It is this formulation that is relevant here, and the corresponding limit $\theta_1 \to 0$ is just as puzzling.

At this point, a number of questions arise, such as: what about the classical case?  Is there some parallel to classical random walks?  To answer this, it is necessary to identify the classical discrete-time process analogous to (\ref{dtqw}).  This is the {\textit{persistent random walk}} \cite{Weiss2002}
\begin{equation}
\begin{array}{l}
p_R(n,\tau+1) = \alpha p_R(n-1,\tau) + \beta p_L(n-1,\tau) \\
p_L(n,\tau+1) = \alpha p_L(n+1,\tau) + \beta p_R(n+1,\tau),
\end{array}
\label{persistent1}
\end{equation}
with $\alpha + \beta = 1$, where $\alpha = \cos^2 \theta$.  This classical process could arise by measuring the coin after each step of the DTQW \cite{Godoy95, Romanelli2004L}.  In general, the persistent random walk differs considerably from the simple diffusion process quantized by (\ref{ctqw}).  For example, it is known that a continuum limit of (\ref{persistent1}) with $\alpha \to 1$ ($\theta \to 0$) \footnote{{The precise limit requires introducing space-time lattice spacings of $\Delta x$ and $\Delta t$.  Then, one lets $\beta = 1-\alpha = v^2 \Delta t /(2 D)$, $\Delta x = v \Delta t$, taking the limit $\Delta t \to 0$ with the continuum coordinates $x = n \Delta x, t = \tau \Delta t$ held constant.}} yields the telegrapher's equation for $p = p_R + p_L$ \cite{Weiss2002,Godoy95}
\begin{equation}
\partial_t p(x,t) = D[\partial_x^2 p(x,t) - v^{-2} \partial_t^2 p(x,t)].
\end{equation}
This evolution has characteristics of both wave propagation and diffusion, becoming the latter only in the limit $v \to \infty$; the relationship with the Dirac equation was explored some time ago \cite{Gaveau84}.

To complete the correspondence with (\ref{ctqw}) and (\ref{dtqw}) requires consideration of the limit $\alpha \to 0$ ($\theta \to \pi/2$), which even classically appears quite unfamiliar.  Nevertheless, one can proceed by iterating (\ref{persistent1}) once to find
\begin{widetext}
\begin{equation}
\begin{array}{l}
p_R(n,\tau+2) = \alpha^2 p_R(n-2,\tau) + \alpha \beta [p_L(n-2,\tau)+p_L(n,\tau)] + \beta^2 p_R(n,\tau) \\
p_L(n,\tau+2) = \alpha^2 p_L(n+2,\tau) + \alpha \beta [p_R(n+2,\tau)+p_R(n,\tau)] + \beta^2 p_L(n,\tau).
\end{array}
\label{persistent2}
\end{equation}
\end{widetext}
Then, letting $t = \tau \Delta t$, $\alpha = 1-\beta = 2 \gamma \Delta t$, the limit $\Delta t \to 0$ leads to the result
\begin{equation}
\begin{array}{l}
\partial_t p_R(n,t) = - 2 \gamma p_R(n,t) + \gamma [ p_L(n-2,t) + p_L(n,t)] \\
\partial_t p_L(n,t) = - 2 \gamma p_L(n,t) + \gamma [ p_R(n+2,t) + p_R(n,t)].
\end{array}
\end{equation}
Finally, defining $p(n,t) = p_R(n,t) + p_L(n-1,t)$, one finds by direct subsitution that
\begin{equation}
\partial_t p(n,t) = \gamma [ p(n+1,t) - 2 p(n,t) + p(n-1,t) ].
\label{persistent3}
\end{equation}
Thus, the continuous-time limit, with $\alpha \to 0$, of (\ref{persistent1}) leads to the discretized diffusion equation (\ref{persistent3}), in complete parallel to the quantum case.  

One might also ask: what about higher-dimensional walks?  Is this limit restricted to one dimension?  Indeed, beyond one dimension, or with general graphs, things might not be so simple.  Consider, for example, the unitary cellular automaton proposed by Bialynicki-Birula \cite{BBirula94}.  Here a four-component amplitude on a three-dimensional lattice $\Psi(n_x,n_y,n_z,\tau)$ has an update rule constructed from conditional displacements and coin rotations just like the DTQW.  Its momentum-space propagation matrix looks quite similar to (\ref{fourier2}):
\begin{equation}
U = e^{-i k_x \sigma_z \otimes \sigma_x} e^{-i k_y \sigma_z \otimes \sigma_y} e^{-i k_z \sigma_z \otimes \sigma_z} e^{-i \theta \sigma_x \otimes I},
\end{equation}
but, due to the fact that operators do not commute, a calculation similar to (\ref{fourier3}) immediately breaks down.  This process depends not only on the directions a particle moves on the lattice, but on the actual sequence of its steps.  A similar effect was seen in a recent quantum walk search algorithm \cite{Ambainis2004a} (in two dimensions), where different coin implementations (step sequences) yielded different results.  

However, by using the slightly more symmetrical sequence of operators
\begin{equation}
\begin{array}{ll}
U = & e^{-i k_x \sigma_z \otimes \sigma_x/2} e^{-i k_y \sigma_z \otimes \sigma_y/2} e^{-i k_z \sigma_z \otimes \sigma_z} \\
& \times e^{-i k_y \sigma_z \otimes \sigma_y/2} e^{-i k_x \sigma_z \otimes \sigma_x/2} e^{-i \theta \sigma_x \otimes I},
\end{array}
\end{equation}
the $\theta \to \pi/2$ limit proceeds as above \footnote{{Specifically, one finds that $U(\tau) \to \exp(-i \Phi) \exp(-i H t)$ where $\Phi = \tau \pi/2$ and $H = - 2\gamma(a \sigma_x \otimes I + \vec{b} \cdot \sigma_y \otimes \vec{\sigma})$, with $a = \cos^2 k_x \cos^2 k_y \cos^2 k_z$, $b_{x} = \sin k_x \cos k_x \cos^2 k_y \cos^2 k_z$, $b_{y} = \cos k_x \sin k_y \cos k_y \cos^2 k_z$, and $b_{z} = \cos k_x \cos k_y \sin k_z \cos k_z$.  The two distinct eigenvalues of $H$ are $\pm 2 \gamma \sqrt{a^2 + b_{x}^2 + b_{y}^2 + b_{z}^2} = \pm 2 \gamma \cos k_x \cos k_y \cos k_z$.  Restricting the evolution to one eigenvalue, and Fourier transforming back yields (\ref{ctqw3d}).}} leading to
\begin{equation}
\begin{array}{l}
i\partial_t \Psi(n_x,n_y,n_z,t)  \\
\qquad = -\frac{\gamma}{4} \sum_{\{d_j = \pm 1\}} \Psi(n_x+d_x,n_y+d_y,n_z+d_z,t),
\end{array}
\label{ctqw3d}
\end{equation}
an obvious generalization of (\ref{ctqw}) to three dimensions.  Thus, the limit found here does generalize to higher-dimensional quantum walks.

In summary, I have found the precise limiting procedure needed to map the discrete-time quantum walk to the continuous-time quantum walk.  This procedure was extended to the classical persistent random walk and diffusion on the lattice, and to higher-dimensional quantum walks.  In all cases one finds the initially counterintuitive result that a process in which a particle moves left and right, flipping its direction with nearly {\textit{unit}} probability per time-step, in the limit of continuous time, yields genuine dynamics.  Classically, one finds diffusion, while quantum mechanically, one finds wave propagation on the lattice.  That this connection remained hidden for so long, and the difficulties encountered in higher dimensions, indicate that there remains a great deal to learn about these simple quantum algorithms.

I sincerely thank A. J. Dragt, P. R. Johnson, and S. Lloyd for helpful discussions and comments.  
\bibliography{qwalk}

\begin{thebibliography}{29}
\expandafter\ifx\csname natexlab\endcsname\relax\def\natexlab#1{#1}\fi
\expandafter\ifx\csname bibnamefont\endcsname\relax
  \def\bibnamefont#1{#1}\fi
\expandafter\ifx\csname bibfnamefont\endcsname\relax
  \def\bibfnamefont#1{#1}\fi
\expandafter\ifx\csname citenamefont\endcsname\relax
  \def\citenamefont#1{#1}\fi
\expandafter\ifx\csname url\endcsname\relax
  \def\url#1{\texttt{#1}}\fi
\expandafter\ifx\csname urlprefix\endcsname\relax\def\urlprefix{URL }\fi
\providecommand{\bibinfo}[2]{#2}
\providecommand{\eprint}[2][]{\url{#2}}

\bibitem[{\citenamefont{Farhi and Gutmann}(1998)}]{Farhi98}
\bibinfo{author}{\bibfnamefont{E.}~\bibnamefont{Farhi}} \bibnamefont{and}
  \bibinfo{author}{\bibfnamefont{S.}~\bibnamefont{Gutmann}},
  \bibinfo{journal}{Phys. Rev. A} \textbf{\bibinfo{volume}{58}},
  \bibinfo{pages}{915} (\bibinfo{year}{1998}).

\bibitem[{\citenamefont{Childs et~al.}(2003)\citenamefont{Childs, Cleve,
  Deotto, Farhi, Gutmann, and Spielman}}]{Childs2003}
\bibinfo{author}{\bibfnamefont{A.~M.} \bibnamefont{Childs}},
  \bibinfo{author}{\bibfnamefont{R.}~\bibnamefont{Cleve}},
  \bibinfo{author}{\bibfnamefont{E.}~\bibnamefont{Deotto}},
  \bibinfo{author}{\bibfnamefont{E.}~\bibnamefont{Farhi}},
  \bibinfo{author}{\bibfnamefont{S.}~\bibnamefont{Gutmann}}, \bibnamefont{and}
  \bibinfo{author}{\bibfnamefont{D.~A.} \bibnamefont{Spielman}},
  \emph{\bibinfo{booktitle}{Proceedings of the Thirty-Fifth Annual ACM
  Symposium on Theory of Computing}} (\bibinfo{publisher}{ACM Press},
  \bibinfo{address}{New York}, \bibinfo{year}{2003}), p.
  \bibinfo{pages}{59}.

\bibitem[{\citenamefont{Childs et~al.}(2002)\citenamefont{Childs, Farhi, and
  Gutmann}}]{Childs2002}
\bibinfo{author}{\bibfnamefont{A.~M.} \bibnamefont{Childs}},
  \bibinfo{author}{\bibfnamefont{E.}~\bibnamefont{Farhi}}, \bibnamefont{and}
  \bibinfo{author}{\bibfnamefont{S.}~\bibnamefont{Gutmann}},
  \bibinfo{journal}{Quantum Inf. Process.} \textbf{\bibinfo{volume}{1}},
  \bibinfo{pages}{35} (\bibinfo{year}{2002}).

\bibitem[{\citenamefont{Aharonov et~al.}(1993)\citenamefont{Aharonov,
  Davidovich, and Zagury}}]{Aharonov93}
\bibinfo{author}{\bibfnamefont{Y.}~\bibnamefont{Aharonov}},
  \bibinfo{author}{\bibfnamefont{L.}~\bibnamefont{Davidovich}},
  \bibnamefont{and} \bibinfo{author}{\bibfnamefont{N.}~\bibnamefont{Zagury}},
  \bibinfo{journal}{Phys. Rev. A} \textbf{\bibinfo{volume}{48}},
  \bibinfo{pages}{1687} (\bibinfo{year}{1993}).

\bibitem[{\citenamefont{Meyer}(1996)}]{Meyer96}
\bibinfo{author}{\bibfnamefont{D.~A.} \bibnamefont{Meyer}},
  \bibinfo{journal}{J. Stat. Phys.} \textbf{\bibinfo{volume}{85}},
  \bibinfo{pages}{551} (\bibinfo{year}{1996}).

\bibitem[{\citenamefont{Brun et~al.}(2003)\citenamefont{Brun, Carteret, and
  Ambainis}}]{Brun2003}
\bibinfo{author}{\bibfnamefont{T.~A.} \bibnamefont{Brun}},
  \bibinfo{author}{\bibfnamefont{H.~A.} \bibnamefont{Carteret}},
  \bibnamefont{and} \bibinfo{author}{\bibfnamefont{A.}~\bibnamefont{Ambainis}},
  \bibinfo{journal}{Phys. Rev. Lett.} \textbf{\bibinfo{volume}{91}},
  \bibinfo{pages}{130602} (\bibinfo{year}{2003}).

\bibitem[{\citenamefont{W{\'{o}}jcik and Dorfman}(2003)}]{Wojcik2003}
\bibinfo{author}{\bibfnamefont{D.~K.} \bibnamefont{W{\'{o}}jcik}}
  \bibnamefont{and} \bibinfo{author}{\bibfnamefont{J.~R.}
  \bibnamefont{Dorfman}}, \bibinfo{journal}{Phys. Rev. Lett.}
  \textbf{\bibinfo{volume}{90}}, \bibinfo{pages}{230602}
  (\bibinfo{year}{2003}).

\bibitem[{\citenamefont{Ribeiro et~al.}(2004)\citenamefont{Ribeiro, Milman, and
  Mosseri}}]{Ribeiro2004}
\bibinfo{author}{\bibfnamefont{P.}~\bibnamefont{Ribeiro}},
  \bibinfo{author}{\bibfnamefont{P.}~\bibnamefont{Milman}}, \bibnamefont{and}
  \bibinfo{author}{\bibfnamefont{R.}~\bibnamefont{Mosseri}},
  \bibinfo{journal}{Phys. Rev. Lett.} \textbf{\bibinfo{volume}{93}},
  \bibinfo{pages}{190503} (\bibinfo{year}{2004}); \bibinfo{author}{\bibfnamefont{A.}~\bibnamefont{W{\'{o}}jcik}}
  \bibnamefont{{\it{et~al.}}}, \bibinfo{journal}{Phys. Rev. Lett.}
  \textbf{\bibinfo{volume}{93}}, \bibinfo{pages}{180601}
  (\bibinfo{year}{2004}).

\bibitem[{\citenamefont{Knight et~al.}(2003)\citenamefont{Knight, Rold{\'a}n,
  and Sipe}}]{Knight2003}
\bibinfo{author}{\bibfnamefont{P.~L.} \bibnamefont{Knight}},
  \bibinfo{author}{\bibfnamefont{E.}~\bibnamefont{Rold{\'a}n}},
  \bibnamefont{and} \bibinfo{author}{\bibfnamefont{J.~E.} \bibnamefont{Sipe}},
  \bibinfo{journal}{Phys. Rev. A} \textbf{\bibinfo{volume}{68}},
  \bibinfo{pages}{020301} (\bibinfo{year}{2003}).

\bibitem[{\citenamefont{Blanchard and Hongler}(2004)}]{Blanchard2004}
\bibinfo{author}{\bibfnamefont{P.}~\bibnamefont{Blanchard}} \bibnamefont{and}
  \bibinfo{author}{\bibfnamefont{M.}~\bibnamefont{Hongler}},
  \bibinfo{journal}{Phys. Rev. Lett.} \textbf{\bibinfo{volume}{92}},
  \bibinfo{pages}{120601} (\bibinfo{year}{2004}).

\bibitem[{\citenamefont{Strauch}(2006)}]{Strauch2006}
\bibinfo{author}{\bibfnamefont{F.~W.} \bibnamefont{Strauch}},
  \bibinfo{journal}{Phys. Rev. A} \textbf{\bibinfo{volume}{73}},
  \bibinfo{pages}{054302} (\bibinfo{year}{2006}).

\bibitem[{\citenamefont{Patel et~al.}(2005)\citenamefont{Patel, Raghunathan,
  and Rungta}}]{Patel2005}
\bibinfo{author}{\bibfnamefont{A.}~\bibnamefont{Patel}},
  \bibinfo{author}{\bibfnamefont{K.~S.} \bibnamefont{Raghunathan}},
  \bibnamefont{and} \bibinfo{author}{\bibfnamefont{P.}~\bibnamefont{Rungta}},
  \bibinfo{journal}{Phys. Rev. A} \textbf{\bibinfo{volume}{71}},
  \bibinfo{pages}{032347} (\bibinfo{year}{2005}).

\bibitem[{\citenamefont{Romanelli et~al.}(2004)\citenamefont{Romanelli,
  {Sicardi Shifino}, Siri, Abal, Auyuanet, and Donangelo}}]{Romanelli2004L}
\bibinfo{author}{\bibfnamefont{A.}~\bibnamefont{Romanelli}},
  \bibinfo{author}{\bibfnamefont{A.~C.} \bibnamefont{{Sicardi Shifino}}},
  \bibinfo{author}{\bibfnamefont{R.}~\bibnamefont{Siri}},
  \bibinfo{author}{\bibfnamefont{G.}~\bibnamefont{Abal}},
  \bibinfo{author}{\bibfnamefont{A.}~\bibnamefont{Auyuanet}}, \bibnamefont{and}
  \bibinfo{author}{\bibfnamefont{R.}~\bibnamefont{Donangelo}},
  \bibinfo{journal}{Physica A} \textbf{\bibinfo{volume}{338}},
  \bibinfo{pages}{395} (\bibinfo{year}{2004}).

\bibitem[{\citenamefont{Konno}(2002)}]{Konno2002}
\bibinfo{author}{\bibfnamefont{N.}~\bibnamefont{Konno}},
  \bibinfo{journal}{Quantum Inf. Process.} \textbf{\bibinfo{volume}{1}},
  \bibinfo{pages}{345} (\bibinfo{year}{2002});
\bibinfo{author}{\bibfnamefont{G.}~\bibnamefont{Grimmett}},
  \bibinfo{author}{\bibfnamefont{S.}~\bibnamefont{Janson}}, \bibnamefont{and}
  \bibinfo{author}{\bibfnamefont{P.~F.} \bibnamefont{Scudo}},
  \bibinfo{journal}{Phys. Rev. E} \textbf{\bibinfo{volume}{69}},
  \bibinfo{pages}{026119} (\bibinfo{year}{2004}); 
\bibinfo{author}{\bibfnamefont{A.~D.} \bibnamefont{Gottlieb}},
  \bibinfo{journal}{Phys. Rev. E} \textbf{\bibinfo{volume}{72}},
  \bibinfo{pages}{047102} (\bibinfo{year}{2005});
\bibinfo{author}{\bibfnamefont{N.}~\bibnamefont{Konno}},
  \bibinfo{journal}{Phys. Rev. E} \textbf{\bibinfo{volume}{72}},
  \bibinfo{pages}{026113} (\bibinfo{year}{2005}).

\bibitem[{\citenamefont{Schweber}(1986)}]{Schweber86}
\bibinfo{author}{\bibfnamefont{S.~S.} \bibnamefont{Schweber}},
  \bibinfo{journal}{Rev. Mod. Phys.} \textbf{\bibinfo{volume}{58}},
  \bibinfo{pages}{449} (\bibinfo{year}{1986});
\bibinfo{author}{\bibfnamefont{R.~P.} \bibnamefont{Feynman}} \bibnamefont{and}
  \bibinfo{author}{\bibfnamefont{A.~R.} \bibnamefont{Hibbs}},
  \emph{\bibinfo{title}{Quantum Mechanics and Path Integrals}}
  (\bibinfo{publisher}{McGraw-Hill, New York}, \bibinfo{year}{1965}).

\bibitem[{\citenamefont{Richardson}(1991)}]{Richardson91}
\bibinfo{author}{\bibfnamefont{J.~L.} \bibnamefont{Richardson}},
  \bibinfo{journal}{Comp. Phys. Comm.} \textbf{\bibinfo{volume}{63}},
  \bibinfo{pages}{84} (\bibinfo{year}{1991}).

\bibitem[{\citenamefont{Boghosian and {Taylor IV}}(1998)}]{Boghosian98b}
\bibinfo{author}{\bibfnamefont{B.~M.} \bibnamefont{Boghosian}}
  \bibnamefont{and} \bibinfo{author}{\bibfnamefont{W.}~\bibnamefont{{Taylor
  IV}}}, \bibinfo{journal}{Physica D} \textbf{\bibinfo{volume}{120}},
  \bibinfo{pages}{30} (\bibinfo{year}{1998}).

\bibitem[{\citenamefont{Weiss}(2002)}]{Weiss2002}
\bibinfo{author}{\bibfnamefont{G.~H.} \bibnamefont{Weiss}},
  \bibinfo{journal}{Physica A} \textbf{\bibinfo{volume}{311}},
  \bibinfo{pages}{381} (\bibinfo{year}{2002}).

\bibitem[{\citenamefont{Godoy and Espinosa}(1995)}]{Godoy95}
\bibinfo{author}{\bibfnamefont{S.}~\bibnamefont{Godoy}} \bibnamefont{and}
  \bibinfo{author}{\bibfnamefont{F.}~\bibnamefont{Espinosa}},
  \bibinfo{journal}{Phys. Rev. E} \textbf{\bibinfo{volume}{52}},
  \bibinfo{pages}{3381} (\bibinfo{year}{1995}).

\bibitem[{\citenamefont{Gaveau et~al.}(1984)\citenamefont{Gaveau, Jacobson,
  Kac, and Schulman}}]{Gaveau84}
\bibinfo{author}{\bibfnamefont{B.}~\bibnamefont{Gaveau}},
  \bibinfo{author}{\bibfnamefont{T.}~\bibnamefont{Jacobson}},
  \bibinfo{author}{\bibfnamefont{M.}~\bibnamefont{Kac}}, \bibnamefont{and}
  \bibinfo{author}{\bibfnamefont{L.~S.} \bibnamefont{Schulman}},
  \bibinfo{journal}{Phys. Rev. Lett.} \textbf{\bibinfo{volume}{53}},
  \bibinfo{pages}{419} (\bibinfo{year}{1984}).

\bibitem[{\citenamefont{Bialynicki-Birula}(1994)}]{BBirula94}
\bibinfo{author}{\bibfnamefont{I.}~\bibnamefont{Bialynicki-Birula}},
  \bibinfo{journal}{Phys. Rev. D} \textbf{\bibinfo{volume}{49}},
  \bibinfo{pages}{6920} (\bibinfo{year}{1994}).

\bibitem[{\citenamefont{Ambainis et~al.}(2005)\citenamefont{Ambainis, Kempe,
  and Rivosh}}]{Ambainis2004a}
\bibinfo{author}{\bibfnamefont{A.}~\bibnamefont{Ambainis}},
  \bibinfo{author}{\bibfnamefont{J.}~\bibnamefont{Kempe}}, \bibnamefont{and}
  \bibinfo{author}{\bibfnamefont{A.}~\bibnamefont{Rivosh}},
  \emph{\bibinfo{booktitle}{Proceedings of the Sixteenth Annual ACM-SIAM
  Symposium on Discrete Algorithms}} (\bibinfo{publisher}{SIAM Press}, \bibinfo{address}{Philadelphia}, \bibinfo{year}{2005}), p. \bibinfo{pages}{1099}.

\end{thebibliography}
\end{document}